\documentclass[amssymb,amsmath,aps,floatfix,nofootinbib, preprintnumbers ,12pt]{revtex4-1}

\usepackage{epsfig}

\usepackage{color}

\usepackage{graphicx}

\def\beq{\begin{equation}}

\def\eeq{\end{equation}}

\def\be{\begin{equation}}

\def\ee{\end{equation}}

\def\bea{\begin{eqnarray}}

\def\eea{\end{eqnarray}}

\newcommand{\gsim}{\lower.7ex\hbox{$\;\stackrel{\textstyle>}{\sim}\;$}}

\newcommand{\lsim}{\lower.7ex\hbox{$\;\stackrel{\textstyle<}{\sim}\;$}}

\begin{document}

\preprint{IPMU 10-0039}

\preprint{NSF-KITP-10-021}

\bigskip

\title{Entropic Inflation}

\author{Damien A. Easson$^{1,2,3}$}
\email{easson@asu.edu}
\author{Paul H. Frampton$^{1,4}$}
\email{frampton@physics.unc.edu}
\author{George F. Smoot$^{1,5,6,7,8}$}
\email{gfsmoot@lbl.gov}
\affiliation{$^1$Institute for the Physics and Mathematics of the Universe,
University of Tokyo, Kashiwa, Chiba 277-8568, Japan}
\affiliation{$^2$ Department of Physics \& School of Earth and
Space Exploration \& Beyond Center,
Arizona State University, Tempe, AZ 85287-1404, USA}
\affiliation{ $^3$Kavli Institute for Theoretical Physics, University of California, Santa Barbara, CA 93106-4030, USA}
\affiliation{ $^4$Department of Physics and Astronomy,
University of North Carolina, Chapel Hill, NC 27599, USA}
\affiliation{$^5$Lawrence Berkeley National Lab, 1 Cyclotron Road, Berkeley, CA 94720, USA}
\affiliation{$^6$Physics Department, University of California, Berkeley, CA 94720, USA}
\affiliation{$^7$Institute for the Early Universe, Ewha Womans University \& Advanced Academy, Seoul, Korea}
\affiliation{$^8$Chaire Blaise Pascale, Universite Paris Denis Diderot, Paris, France}

\begin{abstract}

One of the major pillars of modern cosmology theory                                                                                                                                                                                                                                                                is a period of  accelerating expansion
in the early universe. This accelerating expansion, or inflation,  must be sustained for at least 30 e--foldings.
One mechanism used to drive the acceleration is the addition of a new energy field, called the Inflaton; 
often this is a scalar field.
We propose an alternative mechanism which, like our approach to explain the late-time accelerating universe, uses the entropy and temperature intrinsic to information holographically stored on a surface enclosing the observed space. 
The acceleration is due in both cases to an emergent entropic force, naturally arising from the information
storage on the horizon.

\end{abstract}

\maketitle

\newpage

\section{Introduction}

\bigskip

\bigskip

\noindent A major idea, incorporated in modern theoretical cosmology, is the 
existence of a period of  accelerating expansion
early in the universe's existence, named Inflation \cite{G,AS,Linde}.
This accelerating expansion must be sustained for at least 30 e--foldings.
To accommodate the early highly accelerated expansion of the universe,
one popular idea is to invoke a scalar Inflaton field and a concomitant
Inflaton potential.

\bigskip

\noindent Many observations, particularly the angular spectrum of
the Cosmic Microwave Background (CMB) anisotropies and polarization, the 
power spectrum of density perturbations observed for the Large Scale Structure (LSS),
and other observations such as the flatness, isotropy, homogeneity, and other
features all support  the case for Inflation; although, the detailed dynamics
underlying inflation remain ambiguous, mainly because the evidence
for inflation, while impressive, is circumstantial.
Direct evidence for inflation is hard to come by, and may have to await
detection of gravitational waves induced thereby. Since no
gravitational waves of any type have been directly detected
(only indirectly from binary pulsars), such confirmation may
not be forthcoming in the very near future.

\bigskip

\noindent Nevertheless, despite these caveats, we shall accept the idea 
of an early extended period of acceleration
as a solid component of cosmology theory and in the present paper
conjecture that an entropic force~\cite{EFS1} can be responsible for, 
not only the present accelerating expansion, but the inflationary epoch as well. 
This acts to bolster one's confidence
that the entropic approach is more economical and convenient,
although it may be generally equivalent to the alternative 
explanation by quantum field theory.

\noindent \section{Standard Inflaton}

\bigskip

\noindent On the basis of general relativity theory, together
with the cosmological principle of homogeneity and isotropy,
the scale factor $a(t)$ in the resulting FRW metric satisfies \cite{F,L}
the Friedmann-Lema\^{i}tre equation

\begin{equation}
H(t)^2 = \left( \frac{\dot{a}}{a} \right)^2 = \left( \frac{8 \pi G}{3} \right) \rho 
\label{FLequation}
\end{equation}

\bigskip

\noindent where $\rho$ is the sum of the energy density sources, 
which drive the expansion of the universe. Two established contributions to $\rho$
are $\rho_m$ from matter (including dark matter) and $\rho_{\gamma}$
radiation, so that

\begin{equation}
\rho \supseteq \rho_{m} + \rho_{\gamma}
\label{rho}
\end{equation}

\noindent with  $\rho_{m}(t)  =  \rho_{m}(t_0) a(t)^{-3}$
and $\rho_{\gamma} (t)  = \rho_{\gamma}(t_0) a(t)^{-4}$.

\bigskip

\noindent To produce the accelerated expansion of inflation, a  popular
approach is to add to the sources in Eq.(\ref{FLequation}) an inflaton term $\rho_{I}(t)$ 
with nearly constant density

\begin{equation}
\rho_{I} (t) \equiv \rho_{I}
\,.
\label{rhoI}
\end{equation}

\bigskip

\noindent Matter is modeled as a fluid with equation of state $p = \omega \rho$.
For most models of inflation, the potential energy far outweighs the kinetic term and it approaches the
case $\omega = -1$ (as it does for a cosmological constant, $\Lambda$).
Discarding the matter and radiation terms which are negligible during inflation
we can easily integrate the Friedmann-Lema\^{i}tre equation to find

\begin{equation}
a(t) = a(t_0) ~ e^{ H t}
\label{CC}
\end{equation}

\bigskip

\noindent where $\Lambda \equiv 8 \pi G \rho_{I}$ and $H = \sqrt{ c^2 \Lambda / 3}$.

\noindent
Quantum mechanical fluctuations of the inflaton field are made real with an amplitude set by the expansion rate
and by the equation of state.
In the limiting de Sitter case one would anticipate a scale invariant spectrum of density perturbations.
Slow-roll inflation is observationally preferred resulting in a slight tilt of the perturbations away from scale invariance 
as functions of the derivatives of the inflaton potential.

\bigskip

\noindent  With this background, we now propose a different viewpoint for 
inflation which provides new perspectives on the underlying physics.

\section{Entropic Inflation}

\noindent We previously found the Friedmann-Lema\^{i}tre equation
motivated by entropic force and an effective
surface term.
The entropic force acceleration equation was

\begin{equation}
\frac{\ddot a}{ a} = - \frac{4 \pi G}{ 3} \left(\rho + \frac{3 P}{c^2} \right) + H^2
\label{good}
\end{equation}

\noindent This is remarkably like the surface term order of magnitude estimate presented in \cite{EFS1} except for a $3/2 \pi$ factor.
With the Hawking temperature description, the coefficient would have been 1/2.
There is some freedom here and we chose the value that leads to nice equations in the two limiting cases.
It is easy to show that if $H^2$ is highly dominant over the $  \frac{4 \pi G}{ 3} (\rho + 3 P/c^2)$,
the solution to the equation is simply de Sitter  $a(t) = a(t_0) e^{H(t-t_0)}$.
The alternate equation motivated by surface curvature is

\begin{equation}
\frac{\ddot a}{ a} = - \frac{4 \pi G}{ 3} \left(\rho +  \frac{3 P}{c^2} \right) + \frac{3}{2 \pi} H^2  + \frac{3}{4 \pi} \dot H
\label{bad}
\end{equation}
which may work better for fitting the data, or a rigorous derivation, but does not have the simplicity of Eq. (\ref{good}).
We generalized this to the form

\begin{equation}
\frac{\ddot a}{ a} = - \frac{4 \pi G}{ 3} \left(\rho +  \frac{3 P}{c^2} \right) + C_H H^2  + C_{\dot H} \dot H
\label{general}
\end{equation}
where we anticipate the coefficients to be bounded by $  C_H < 1$ and
$0 \leq C_{\dot H} \lsim \frac{3}{4 \pi} $ to be determined by observations.

\subsection{Entropic Force Considerations}

\noindent
In our previous work, we showed the entropic force is directed at the information encoded on the horizon surface.
We now derive a generalized expression for the pressure, showing it is negative, thus providing a tension in the direction of the surface.

\noindent
The semi-classical entropy on the screen, e.g. on the horizon, is

\bigskip

\begin{equation}
S_H = \frac{k_B c^3}{G \hbar} \frac{A}{4} = \frac{k_B c^3} {G \hbar} \pi R_H^2 = \frac{k_B c^3} {G \hbar} \pi \left( \frac{c}{H} \right)^2 \sim (2.6 \pm 0.3) \times 10^{122} k_B
\end{equation}
We modify this formula for the next level corrections motivated by Boltzmann~\cite{Boltzmann}
and by string theory approaches to entropy~\cite{Vafa}
and to inflation~\cite{KKLT,KKLMMT}
to include a term that takes into account the number of ways this information can be stored on the surface.
It has been established for many years in string theory~\cite{Vafa,Sol}, and in quantum loop gravity \cite{Rovelli, Ashtekar, Kaul}, 
that the first order corrected equation should be of the form

\begin{equation}
S = \frac{A}{4 \ell_{Pl}^2} + \varrho ~ \ln \left( \frac{A}{\ell_{Pl}^2} \right) +  \mathcal{O} \left( \frac{\ell_{Pl}^2}{A} \right)
~~~\equiv  \frac{1}{4} \frac{A}{A_{Pl}} +  \varrho ~ \ln \left( \frac{A}{A_{Pl}} \right) +  \mathcal{O} \left( \frac{A_{Pl}}{A} \right)
\end{equation}
and that there are no correction terms stronger than logarithmic dependence, although there is no sharp
consensus on the coefficient $\varrho$. In the above, $A_{Pl} \equiv  \ell_{Pl}^2$ is the Planck area, defined in terms of the Planck length $ \ell_{Pl} =\sqrt{\hbar G/c^3}$.

\bigskip

\noindent One can also go directly to the Boltzmann epitaphal formula $S = k_B \ln W$,
where $W$ is the number of microstates that produce the same macrostate;
in this case the number of ways that the same information can be stored on the surface.
This should be proportional to a factor times the logarithm of the number of bits as the first order correction.

\bigskip

\noindent
One can estimate the number of states $W$ by considering that there are $N = A/4A_{Pl} $ sites for the basic information on the screen and that the number of possible bits of information must be some factor times the number of sites $N$.
Thinking in terms of excited states and anticipating that only the lowest and next energy level will be excited,
one would expect of order 2 bits are available at each site $N$ giving
us $W = 2N! / ( (N+1)!(N!)) = \Gamma(2N)/(\Gamma(N+1) \Gamma(N))$ which is the number of ways that 2 bits can be placed in $N$ sites.

\bigskip

\noindent
This yields
$S = k_B \frac{A}{4A_{Pl}}  \ln(2) + \frac{1}{2}C \ln( A / A_{Pl}) + constant$
from the Stirling's approximation and a correction term of the form $g k_B \ln( A / A_{Pl})$. 
One does not care about the added constant as 
it has no effect in the force/acceleration.

\begin{equation}
S_H = \frac{k_B c^3}{G \hbar} \frac{A}{4} + g k_B \ln{\frac{A}{A_{Pl}}}  =\frac{ k_B}{4} \frac{A}{A_{Pl}} + g k_B \ln{\frac{A}{A_{Pl}}}
\label{semiplus}
\end{equation}

\bigskip

\noindent
where the factor $g$ includes the effective number of independent degrees of freedom, since entropy will be the sum over each degree of freedom and we have included the factor of $k_B$ not shown in the string theory and loop gravity formula.

\bigskip

\noindent Increasing the Hubble radius $R_H = c H^{-1}$, by $d r$, increases the entropy by $d S_H$ according to

\begin{equation}\label{dSH}
\frac{d S_H}{d r}  =  \frac{k_B c^3} {4 G \hbar} \frac{d A}{d r}  + g k_B \frac{1}{A}  \frac{d A}{d r}
= \left( \frac{k_B c^3} {4G \hbar} +  \frac{g k_B}{4 \pi c^2} H^2 \right)   \frac{d A}{d r}
\end{equation}

where in the last expression we have used $A = 4 \pi R_H^2$.

\bigskip

\noindent
This will give us another term in our acceleration equation  Eq. (\ref{good})

\begin{equation}
\frac{\ddot a}{ a} = - \frac{4 \pi G}{ 3}  \left(\rho + \frac{3 P}{c^2} \right)  + H^2 + \kappa H^4 = - \frac{4 \pi G}{ 3}  \left(\rho + \frac{3 P}{c^2} \right) + H^2 \left(1 +  \frac{g}{ \pi} \frac{H^2}{ H_{Pl}^2} \right)
\label{goody}
\end{equation}
where $\kappa = \frac{4 G \hbar}{c^3}    \frac{g} {4 \pi c^2}= \frac{g}{ \pi H_{Pl}^2} $, and we have defined $H_{Pl} = \ell_{Pl}^{-1} c$.

\bigskip

\noindent The entropic force is

\begin{equation}\label{forcer}
F_r = -  \frac{dE}{dr} = -  T  \frac{dS}{dr}  = - T_\beta \frac{ d S_H}{d r}
\end{equation}
where we identified the crucial temperature $T_\beta$, the entropic temperature
of the universe, precisely as in our earlier work  as
\footnote{Note that it is important that $T_{\beta}$ is non-zero, otherwise
the acceleration would vanish.}

\begin{equation}
T_{\beta} = \frac{\hbar}{  k_B }~\frac{H }{ 2 \pi} \sim 3 \times 10^{-30} K .
\label{T-beta}
\end{equation}
Using (\ref{dSH}), (\ref{forcer}) and (\ref{T-beta}) ,

\begin{equation}
F_r =- \frac{c^4 }{ G}   \left(1 +  \frac{g}{\pi} \frac{H^2}{ H_{Pl}^2} \right)
\end{equation}
where the minus sign indicates the force is pointing in the direction of increasing entropy or the screen, which in this case is the Hubble horizon.

\bigskip

\noindent

The pressure exerted follows directly from the entropic force as

\begin{equation}
P_r =  \frac{F_r}{A}   = -  \frac{1}{A} \frac{c^4 }{ G}  \left(1 +  \frac{g}{ \pi} \frac{H^2}{ H_{Pl}^2} \right)
= -  \frac{ c^2 H^2}{4 \pi G}   \left(1 +  \frac{g}{ \pi} \frac{H^2}{ H_{Pl}^2} \right) 
= -  \frac{2}{3} \rho_{c} c^2   \left(1 +  \frac{g}{ \pi} \frac{H^2}{ H_{Pl}^2} \right)
\end{equation}
where $\rho_c$ is the critical energy density given by $\rho_c = \frac{3H^2}{8 \pi G}$.
Hence, the expression for the entropic pressure is modified to

\begin{equation}
P_r =   -  \frac{2}{3} \rho_{c} c^2 \left( 1 + \frac{g}{ \pi} \frac{ H^2}{H_{Pl}^2} \right) 
= -  \frac{2}{3} \rho_{c0} c^2 \left( \frac{H^2}{H_0^2} + \frac{g}{ \pi} \frac{ H^4}{H_0^2 H_{Pl}^2} \right) 
\end{equation}

\bigskip

\noindent At late times, with small $H$, this is of course,  close to the value of the currently measured dark energy / cosmological constant negative pressure (equals tension).
In this case the tension does not arrive from the negative pressure of dark energy but from the entropic tension (outward pointing pressure) due to the (information) entropy content of the surface.
This is equivalent to the outward acceleration $a_H = cH$, attained by substituting $T_\beta$ into the Unruh relation $a_H = 2 \pi c k_B T/\hbar$.

\bigskip

\noindent However, and this is the main result of the present paper, in the very early universe when $H$ was large this can be a significant correction.
Surprisingly, and delightfully, we will see it can lead to the driving force behind inflation.

\bigskip

\section{Inflation with the Entropy Correction Term}


\noindent
We now consider the case that tends toward an accelerating universe which occurs when the scale factor $a$ is very small and the expansion rate is near exponential.
This may well be the case in the very early universe.

\noindent
Starting with the generalized acceleration Eq. (\ref{goody}) and using the identity 

\begin{equation}
 \frac{\ddot a}{ a}  = \dot H + H^2
\end{equation}
 we are lead to the standard GR equation for $\dot H$  {\it independent of $C_H$ and $ C_{\dot H} $} \rm with only an extra term from the logarithmic correction to the entropy:

\begin{equation}
 \dot H = -4 \pi G \left(\rho + \frac{ P }{ c^2}\right) +  \frac{g}{ \pi} \frac{H^4}{ H_{Pl}^2}
\end{equation}

\noindent 
One approaches a de Sitter - an accelerating universe with $\dot H = 0$, that is an unchanging accelerating expansion rate and $a(t) = a(t_0) e^{H(t - t_0)}$ scale factor, when

\begin{equation}
 \frac{g}{ \pi} \frac{H^4}{ H_{Pl}^2}   =  4 \pi G \left(\rho + \frac{ P}{c^2} \right) =  4 \pi G \rho (1 + w)  \rightarrow \frac{16  \pi G}{ 3} \rho
\label{deSit}
\end{equation}
where the last holds at relativistically high energies so that $P = \rho c^2 /3$. Thus, to obtain a sufficiently long-lived radiation dominated
period may require tuning.

\bigskip

\noindent
Now we must be careful not to use our simple scaling laws for the matter and radiation content but to express them generally as we do not know how many degrees of freedom are available at very high energies (temperatures).
As long as we are in the regime where the expansion rate $H$ is both large and essentially constant, the density in the universe is constant rather than scaling as radiation, $1/a(t)^4$.
It behaves very much like a cosmological constant with its density independent of the comoving volume.

\bigskip

\noindent
In this model it is the thermal radiation from the temperature of the horizon  $T$ that heats the interior space and provides the energy density $ \rho = G(T) u(T) /c^2 = G(T) \frac{4  \sigma}{c^3} T^4$ where $T = \hbar H / 2 \pi k_B$ is the temperature of the patch
and $G(T)$ is the effective number of degrees of freedom at temperature $T$.
Substituting in the horizon temperature we find

\begin{equation}
 \frac{g}{ \pi} \frac{H^4}{ H_{Pl}^2}   =  4 \pi G \left(\rho + \frac{ P}{c^2} \right) =  4 \pi G \rho (1 + w)   \rightarrow  \frac{16 \pi G}{ 3 c^3} G(T) 4 \sigma T_H^4 =  \frac{16 \pi G c^3}{ 3 c^3} G(T) 4 \sigma \left(  \frac{\hbar H }{ 2 \pi k_B}  \right)^4
\label{deSitt}
\end{equation}
 which becomes

\begin{eqnarray}
 \frac{g}{ \pi H_{Pl}^2}   & = & \frac{16 \pi G}{ 3 } G(T) 4 \sigma \left(  \frac{\hbar  }{ 2 \pi k_B}  \right)^4 =  \frac{16 \pi G}{ 3c^3} G(T) 4 \frac{ \pi^2 k_B^4}{60 \hbar^3 c^3} \left(  \frac{\hbar  }{ 2 \pi k_B}  \right)^4  \nonumber \\
 &=&  G(T) \frac{G \hbar}{45 \pi c^5} = G(T) \frac{A_{Pl} }{45 \pi c^2} = \frac{G(T)}{45 \pi H_{Pl}^2}
\label{deSitter}
\end{eqnarray}

\bigskip

\noindent
Thus

\begin{equation}
g =  \frac{G(T)}{45}   ~~~~{\rm or} ~~~~ G(T) = 45 g
\end{equation}

\bigskip

\noindent
This tells us about our correction to the semi-classical entropy in Eq. (\ref{semiplus})

\begin{equation}
S_H = \frac{k_B c^3}{G \hbar} \frac{A}{4} + g k_B \ln\left(\frac{A}{A_{Pl}}\right) = \frac{k_B}{4} \frac{A}{A_{Pl}} + \frac{G(T_*)}{45} k_B ~\ln\left(\frac{A}{A_{Pl}}\right)
\label{semiplusc}
\end{equation}
where $G(T_*)$ is the effective number of degrees of freedom, at the appropriate temperature $T_*$ which is either something such as the temperature that gives the number of degrees of freedom associated with the substructure of spacetime (e.g. strings), the effective number of degrees of freedom during inflation, or the effective number of degrees of freedom in the universe at the present epoch.
The equality was derived from the early high accelerating rate of expansion epoch.
We would anticipate that entropy is additive and that in equilibrium each degree of freedom should add an equal amount to the total entropy.

\bigskip

\noindent
If this is satisfied, then the universe will continue to accelerate its expansion in
de Sitter mode because the logarithmic correction in Eq. (\ref{semiplusc}) 
is suffucient for this purpose.

\bigskip

\noindent 
The entropic viewpoint does not yet shed new light on how
to exit the inflationary epoch. In the quantum field theory approach, we expect
that the exit will be by quantum fluctuations in the inflaton field. We have deliberately
kept our discussion at the semi-classical
level because the relationship between entropic
gravity and quantum mechanics, if any, remains obscure.
Entropic gravity can be viewed as demoting gravity from a fundamental interaction
and hence demoting the graviton from an elementary particle status to be more like a phonon than a photon.
Nevertheless, it is possible that an effective inflaton and inflaton potential might
be derivable from entropic gravity to understand better the relationship between the two descriptions.

\bigskip

\noindent
Naturally, 
some regions will have a slight over density and slow the expansion rate and lower the temperature.
If the number of degrees of freedom is an inverse function of temperature, and hence increases
 as the temperature decreases from highest temperatures,
e.g. due to extra degrees of freedom appearing in complex condensed systems,
then the thermal energy density of these degrees of freedom will dominate  
over the acceleration and quickly bring the inflationary period to an end.

\bigskip

\noindent
However, if the number of degrees of freedom stays constant, the positive fluctuations  will continue,  slowly decelerating until particles begin to freeze out of thermal equilibrium and leave a residual density higher than the thermal background and one soon switches to a fully decelerating universe.
Here we have to track the $G(T)$ for the matter and energy density in the lower temperature universe with the standard scaling laws proportional to: $1/a(t)^3$, $1/a(t)^4$, for matter and radiation (relativistic species), respectively.
\footnote{
Including the correction energy term in the equation gives nearly the same constraint on $g$ and there is acceleration from the Planck scale down until the acceleration automatically shuts down at roughly $10^{-6} H_{Pl}$. 
At much later epochs when the matter and energy density are sufficiently diluted the acceleration resumes. 
This early slow down may bring the perturbation spectral index down slightly.}
We can reasonably expect that this also produces the near scale invariant density perturbation spectrum, and leave a detailed study of this topic for future discussion.

\bigskip

\noindent
Assuming that we have density perturbation of the level $(\delta \rho / \rho) \leq   10^{-4}$, then it is straightforward to show that for a positive fluctuation, the Hubble expansion rate will follow

\begin{equation}
H = \frac{H_I}{1 + \frac{\delta \rho}{\rho} H_I (t - t_0)}
\end{equation}
So that with the small $\delta \rho / \rho $ we have no issue with getting to order $\rho / \delta \rho \geq 10^4$ e--foldings.

\noindent
If $H_I$ were at the Planck scale, then $\delta \rho / \rho \sim 1 $.
 So we recognize that we need to be at least 3 orders of magnitude below the Planck scale $H_{Pl}$.
 This is the same as for  standard inflationary models.
 We also know if the expansion rate and temperature is at that scale, we should have detected gravitational waves from the metric perturbations, and they are down at least another order of magnitude.

\noindent
We see that $H$ changes by of order $\delta \rho / \rho$ per e-folding.
Thus the deviation of the spectral index from scale invariant should be
$1 - n \sim \delta \rho / \rho$.
This gives a near-scale invariant perturbation spectrum which may
risk marginal disagreement with WMAP7 \cite{Komatsu}.

\section{Discussion}

 \bigskip

 \noindent 
 The inclusion of the first order correction to the semi-classical horizon entropy
 provides a natural source of inflation - accelerated expansion of the early universe.
 This is independent of our coefficients for the late time acceleration
parameterizing entropic terms introduced in \cite{EFS1} to replace dark energy for late time acceleration.

\bigskip

\noindent
The physical mechanism  of an entropic force causation is the same; 
 the difference between the inflationary era and the late-time acceleration lies in that inflation
 comes from the first order correction term to the entropy and the late time acceleration
 comes from the semi-classical terms. 
 The relative sizes of terms in the Friedmann-Lema\^{i}tre equation
are due to the different sizes for the cosmic scale factor and expansion rates.
 This approach provides a hopefully fruitful physical understanding of inflation,
which was lacking in quantum field theory.
 For example we have a constraint on the relationship between the effective number of degrees 
 of freedom of the structures of space-time and the particles and fields that emerge.

\bigskip

\noindent
This entropic viewpoint will hopefully shed light on questions that are challenging 
in a quantum field theoretic inflaton
approach, such as exit from inflation and reheating.
 Surprisingly  we see that string theories could lead to inflation in those circumstances. Observations and future efforts will lead us to better understanding of these phenomena.

\bigskip

 \noindent 
Here inflation, like the late-time accelerated expansion,
is based on the entropic force concept.
Instead of the inflaton, or dark energy, we have the holographic principle and
entropy which combine to dark entropic geometry as the source of the accelerating phases of the universe.

\begin{center}

\section*{Acknowledgements}

\end{center}

\noindent This work was supported by the World Premier
International Research Center Initiative (WPI initiative), MEXT, Japan.
G.F.S. would like to thank Sumit R. Das for simulating discussions on the string 
theory approach to determine the higher order terms for the entropy.
P.H.F. acknowledges useful discussions
with S. Adler and T. Banks at the Gell-Mann festschrift.
The work of D.A.E is supported in part by a Grant-in-Aid for Scientific Research
(21740167) from the Japan Society for Promotion of Science (JSPS), by the DOE, by the Cosmology Initiative at Arizona State University
and by the National Science Foundation (KITP, UCSB) under Grant No. PHY05-51164.
The work of P.H.F. was supported in
part by U.S. Department of Energy Grant No. DE-FG02-05ER41418.
G.F.S. is supported in part by the U.S. Department of Energy under Contract No. DE-AC02-05CH11231, by WCU program of NRF/MEST (R32-2009-000-10130-0),
and by CNRS Chaire Blaise Pascal.

\end{document}